# Decays into $\pi^+\pi^-$ of the $f_0(1370)$ scalar glueball candidate in pp central exclusive production experiments

Ugo Gastaldi[1,2*] and Mirko Berretti[3]

**Abstract**

The existence and properties of the $f_0(1370)$ scalar meson are rather well established from data of antiproton annihilations at rest. However conflicting results from Central Exclusive Production (CEP) experiments of the last millennium and ignorance of data from antiproton annihilations at rest in $H_2$ and $D_2$ bubble chambers have generated doubts on the very existence of the $f_0(1370)$. Properties of $\pi^+\pi^-$ pairs produced in central exclusive production (CEP) reactions observed in old data together with data collected in the current decade at high energy colliders permit to show that $\pi^+\pi^-$ decays of the $f_0(1370)$ meson are directly observable as an isolated peak between 1.1 and 1.6 GeV. Consequences of this observation and prospects for the identification of the scalar glueball ground-state are discussed.

1-CERN, Geneva, Switzerland
2-INFN Sezione di Ferrara, Italy
3-University of Helsinki and Helsinki Institute of Physics, Finland
*Corresponding author, email: gastaldi@cern.ch

## 1-Introduction

There is something special with the $f_0(1370)$ meson. It decays into 4 pions at least six times more frequently than into any two pseudoscalars decay channel. Its dominant $4\pi$ decay mode is via $\sigma\sigma$. This $0^{++}$ meson decays then preferentially into a pair of $0^{++}$ mesons: no angular momentum barrier to be overcome, no violation of generalized Zweig rule if $f_0(1370)$ is prominently a scalar glueball and if also $\sigma$ is essentially made of gluonium.

However not everybody is convinced of the existence of the $f_0(1370)$ as an object distinct from the high energy tail of a $0^{++}$ S-wave $\pi\pi$ continuum which would include the $\sigma$ at low energy [see e.g. refs. 1-5]. Moreover, as will be seen in section 2, there are controversial results from antiproton annihilation experiments at LEAR and central exclusive production experiments at SPS concerning 4 pion decays of $f_0(1370)$ and $f_0(1500)$. The skepticism concerning the very existence of $f_0(1370)$ is supported by the different treatment of $f_0(1370)$ and $f_0(1500)$ in the PDG tables of the last 20 years [6]. The amplitudes of $f_0(1370)$ and $f_0(1500)$ interfere, so it is not possible to establish their properties independently. Yet decay modes of $f_0(1500)$ are given with relative errors sometimes below 15% while decay modes of $f_0(1370)$ are not given, and papers used to establish averages for mass and width of $f_0(1500)$ are discarded for $f_0(1370)$.

At low energies there are many more isoscalar scalars in excess of the two necessary to fill the $0^{++}$ ground state nonet: $\sigma$, $f_0(980)$, $f_0(1370)$, $f_0(1500)$, $f_0(1710)$. The nature of $\sigma$ and of $f_0(980)$ is object of debate [7-13]. The $f_0(980)$ may be a tetraquark or a KKbar object [7,8]. $\sigma$ and $f_0(1370)$ may be respectively the low and high manifestations of the "red dragon" scalar gluonium and $f_0(980)$ and $f_0(1500)$ the two isoscalar members of the $0^{++}$ nonet [1]. If $\sigma$ and $f_0(980)$ were the two isoscalar members of the lowest energy $0^{++}$ nonet two out of the three $f_0$s of highest mass might be radial excitations. If $\sigma$ and $f_0(980)$ could be classified as nonqqbar nor gg objects, $f_0(1370)$, $f_0(1500)$ and $f_0(1710)$ might be the result of the mixing of the two isoscalar members of the $0^{++}$ scalar nonet and of a $0^{++}$ glueball. Hence the importance of establishing firmly the nature of $f_0(1370)$. Indeed there is general consensus that scalar gluonium has not yet been firmly identified. For reviews see [4,5,14-16]. In Ochs's comprehensive review of the status of glueballs [5] there is this warning: "We have not found convincing evidence for $f_0(1370)$ in any of the 2-body decay channels in the study of a large variety of reactions".

In section 2 some relevant data concerning $f_0(1370)$ and $f_0(1500)$ in pbar annihilations at rest [17-26] and in central exclusive production at relatively low energies [27-47] are discussed. In section 3 are recalled the main general features of central exclusive production. In section 4 we discuss data of the STAR experiment at RICH which feature a clear signal of the of $f_0(1370)$ decays into $\pi^+\pi^-$. In section 5 are given a summary of the status of $f_0(1370)$, conclusions and prospects.

## 2- $f_0(1370)$ and $f_0(1500)$ in pbar annihilations at rest and in central production experiments

Data of pbar annihilations at rest provide convincing evidence of the existence of the $f_0(1370)$ meson and permit to determine rather well its properties (mass, width and decay branching ratios). For experimental reviews see [4,14-16].

The $\pi^+\pi^-\pi^+\pi^-$ decays of $f_0(1370)$ were observed already in 1966 in dpbar annihilations at rest (in liquid $D_2$ in the CERN 80 cm bubble chamber) [17]. $0^{++}$ $J^{PC}$ quantum numbers were attributed to the $\pi^+\pi^-\pi^+\pi^-$ structure and $\rho\rho$ dominant decays were suggested. By reanalyzing in 1993 BNL bubble chamber data of dpbar→ $p\pi^-\pi^+\pi^-\pi^+\pi^-$annihilations at rest in liquid $D_2$, Gaspero found that $\sigma\sigma$ is the dominant intermediate state of $f_0(1370) \rightarrow \pi^+\pi^-\pi^+\pi^-$ decays, that interferences of the $\sigma\sigma$ and $\rho\rho$ decay amplitudes were necessary to produce good fits of all the decay angular distributions and mass plots, and established the $0^{++}$ nature of the structure [18](see fig.1 [18]).

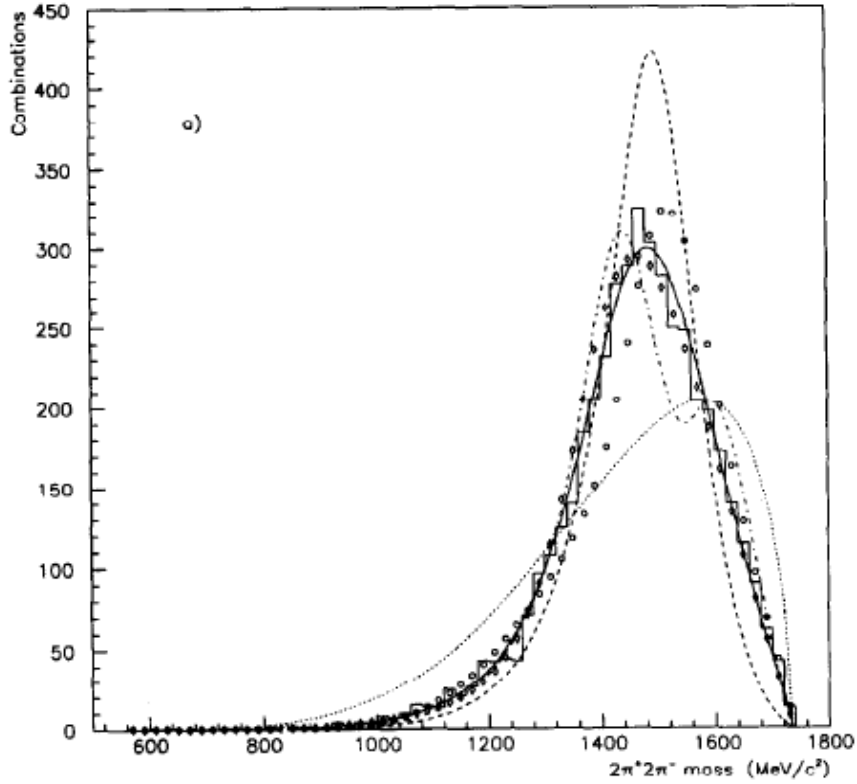

Fig.1: π⁺π⁻π⁺π⁻ invariant mass of dpbar→p π⁻ π⁺π⁻π⁺π⁻ annihilations at rest in liquid $D_2$. The dotted line shows phase space, solid line shows the best fit. Open circles and open diamonds show respectively the predictions of the best fit with only ρρ and σσ intermediate channels (from ref. 18)

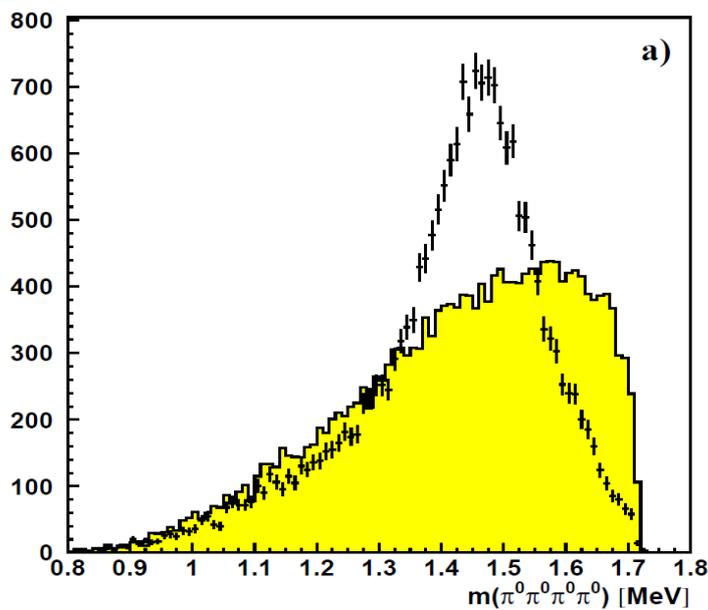

Fig. 2: 4 $π^0$ invariant mass spectrum in dpbar→ pπ⁻4 $π^0$ annihilations. The peak (data points with errors) is generated mainly by $f_0(1370)$ decays into 4$π^0$. The histogram shows phase space (Crystal Barrel data [21]).

At LEAR the Crystal Barrel Collaboration measured pbar annihilations at rest in liquid $H_2$ and $D_2$ with high statistics and excellent gamma detection. An extremely convincing signal of $f_0(1370)$ decays into 4 $\pi^0$ is visible in dpbar $\rightarrow$ p $\pi^-$ $4\pi^0$ data (were no combinatorial background is present) (see fig.2 from [21]) These 4 $\pi^0$ data feature dominant $f_0(1370) \rightarrow \sigma\sigma$ decays, and the $\rho\rho$ decay channel cannot be present, since $\rho \rightarrow 2\pi^0$ decays are forbidden. $f_0(1370)$ decays into $\pi^0\pi^0\pi^+\pi^-$ and $4\pi^0$ in ppbar annihilations at rest in liquid $H_2$ and $\pi^0\pi^0\pi^+\pi^-$ decays in liquid $D_2$ were also measured by CB (see fig. 4). The same set of decay amplitudes into 4 pions gave good fits of all the mass plots (see fig.3 for the $\pi^-4\pi^0$ data set) of all the measured sets of 4 pion annihilation data. The essential results were stable over a period of 5 years of studies [19-22]. The $f_0(1370)$ mass and width turned out to be M=1395+-50 MeV, Γ=275+-50 MeV. The dominant $f_0(1370)$ decay channel is $\sigma\sigma$. It is twice as frequent than the $\rho\rho$ decay channel and much larger than all the allowed and measured $f_0(1370)$ decays into two light mesons ($\pi^+\pi^-$, $\pi^0\pi^0$, $K^+K^-$, $K^0_sK^0_s$, $K^0_lK^0_l$, $\eta\eta$). The $f_0(1500)$ decays into $4\pi$ were studied in the same data: the $f_0(1500)$ mass and width turned out to be compatible with the average values in the PDG's tables: M=1504+-6 MeV, Γ=109+-7 MeV. The $\sigma\sigma$ decays of $f_0(1500)$ are twice more frequent than the $\rho\rho$ decays, but are dominated by the $\pi\pi$ decays [15]. These Crystal Barrel data are not used in PDG's tables to give average mass and width values and decay branching ratios of $f_0(1370)$.

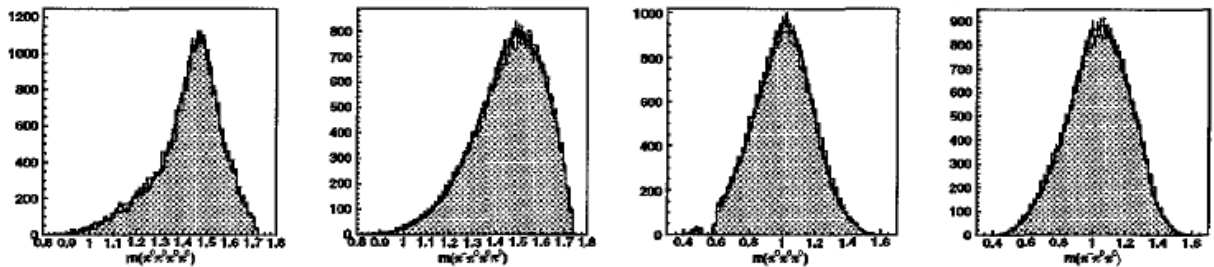

Fig.3 $\pi^-4\pi^0$ data: invariant mass for the combinations (left to right) $4\pi^0$, $\pi^-3\pi^0$, $3\pi^0$, $\pi^-2\pi^0$. Shaded distribution shows fit, points with error bars show data (Crystal Barrel data [20]).

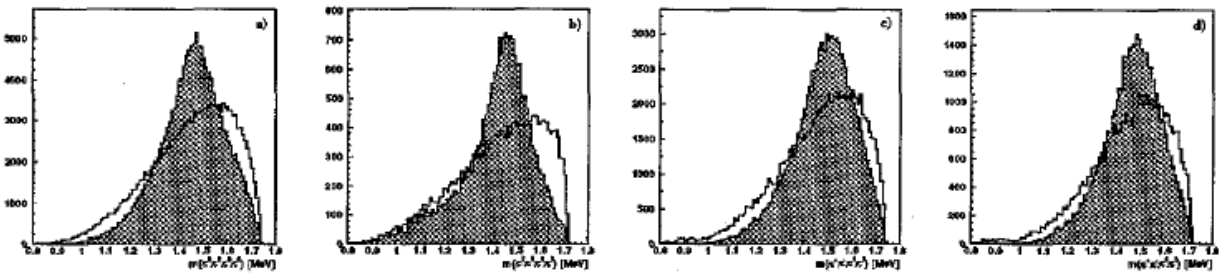

Fig.4: Neutral $4\pi$ invariant mass for all 4 data sets: (a) $5\pi^0$, (b) $\pi^-4\pi^0$, (c) $\pi^+\pi^-3\pi^0$, (d) $\pi^+2\pi^-2\pi^0$. Shaded distribution shows data, curve shows phase space (Crystal Barrel data [20]).

$f_0(1370)$ decays into $K^0_sK^0_s$ are present in data of npbar annihilations at rest in liquid $D_2$ in CERN and BNL bubble chamber data of the sixties [23,24]. These data feature superior topological identification and mass resolution for dpbar $\rightarrow$ p $\pi^-$ $K^0_sK^0_s$ events because of the direct observation and measurement of the annihilation point (end point of the pbar track and starting point of the $\pi^-$ track) and of the two separate vertices of the two $K^0_s$ decays into $\pi^+\pi^-$. The points in the Dalitz plots of the CERN and BNL data were added in a single Dalitz plot in 2000 [25,26] (see fig.2a in [25]).

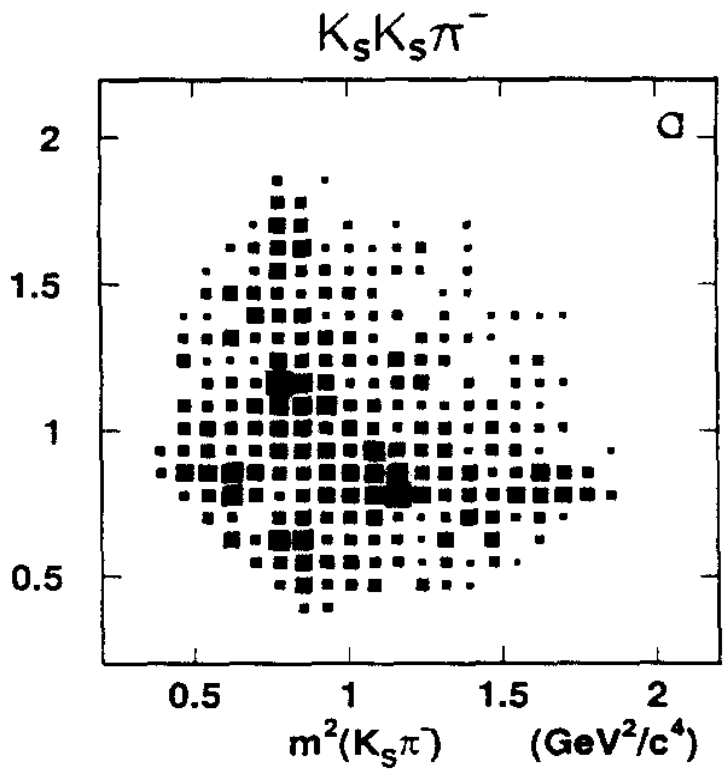

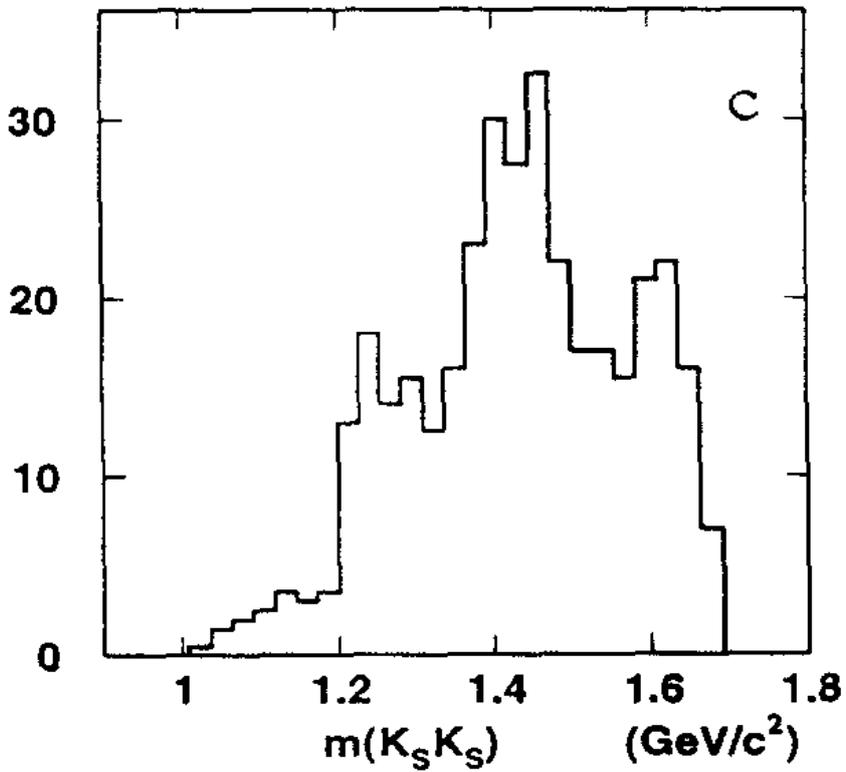

Fig.5 : $K^0_s K^0_s \pi^-$ Dalitz plot (top plate) and $K^0_s K^0_s$ mass plot of dpbar→p $\pi^-$ $K^0_s K^0_s$ annihilations at rest in liquid $D_2$ [25]

The $f_0(1370)$ band and the $K^{-*} \rightarrow K^0_s\pi^-$ bands are clearly visible in the $\pi^-K^0_sK^0_s$ Dalitz plot. The interferences of the $f_0(1370)$ band with the $K^{-*} \rightarrow K^0_s\pi^-$ bands are observable with sharp contrast and witness the resonant nature of the $f_0(1370)$ structure and its limited width. The peak due to the $f_0(1370)$ decays into $K^0_sK^0_s$ dominates the $K^0_sK^0_s$ mass plot (see fig.5). These data, which should answer questions raised in the review of Ochs have long remained unnoticed and are not quoted nor used in available compilations.

Indications of the existence of $f_0(1370)$ were present in the central exclusive production data of pp $\rightarrow$ p$\pi^+\pi^-$p collected at the CERN ISR by the AFS [27,28] and SFM [29,30] spectrometers. AFS and SFM featured different kinematical coverages of the scattered protons. The four-momentum transfer I–tI acceptance window of AFS at $\sqrt{s}$= 63 GeV was 0.01 < I-t I < 0.06 GeV$^2$ at both proton vertices, the $\pi^+\pi^-$ mass spectrum (see fig. 6 left plate) featured a dominant σ signal peaking at about 0.5 GeV followed by an order of magnitude sharp drop of intensity between 0.9 and 1.1 GeV around the $f_0(980)$ mass. Between 1.1 and 1.45 GeV the spectrum featured a broad enhancement (by about 50% ) followed by a second drop of about one order of magnitude moving from 1.4 to 1.6 GeV. The angular distributions featured pure S-wave up to 1.1GeV and dominant S-wave with presence of D-wave between 1.1 and 1.6 GeV. The D-wave component was attributed to a contribution of $f_2(1270)$ not immediately noticeable in the mass spectrum. The S-wave signal in the 1.1-1.5 GeV window was interpreted as the possible manifestation of the $f_0(1370)$ (at those times named $f_0(1400)$). The sharp drop around 1 GeV was interpreted as the result of the interplay of the amplitudes of the high energy tail of the σ (or of the $0^{++}$ continuum) and of the $f_0(980)$, plus the effect of crossing of the KKbar threshold. The drop centered at 1.5 GeV has been associated later on to the interference of the amplitudes of the $f_0(1500)$ with the tail of the continuum, or with the $f_0(1370)$, if the $f_0(1370)$ was indeed the main contributor to the region 1.1-1.6 GeV.

The minimum of the four momentum acceptance of the SFM experiment I-$t_{minSFM}$I > 0.08 GeV$^2$ for both protons was much higher than that of the AFS one: I-$t_{minAFS}$I > 0.01 GeV$^2$. The SFM $\pi^+\pi^-$ mass plot measured at $\sqrt{s}$ =62 GeV (see fig. 6 right plate) is reminiscent of the AFS one, but with marked differences. The signal raises nearly smoothly from threshold to 0.9 GeV and does not feature the broad σ signal which peaks at 0.5 GeV in the AFS spectrum. Across 1 GeV a sharp drop occurs similar to the AFS one. However the signal lowers in the SFM case by a factor 2 and not by a factor of about 10 as in AFS. A clear peak appears in the SFM spectrum centered at about 1.25 GeV followed by a drop by a factor of about 5 moving from 1.1 to 1.5 GeV. The peak was attributed to the $f_2(1270)$ because its signal is directly visible in the mass plot and is confirmed by the D-wave component in the partial wave analysis.

The AFS and SFM data alone do not permit to discriminate between the hypothesis of a broad continuum peaking at about 0.5 GeV and decaying down beyond 1.6 GeV with two drastic interferences in correspondence of the $f_0(980)$ and $f_0(1500)$ mesons, and the alternative hypothesis of the $f_0(1370)$ resident between the $f_0(980)$ and the $f_0(1500)$ . Therefore the AFS and SFM data alone cannot be used to settle the question of the existence of $f_0(1370)$ in central exclusive processes. By the way, these AFS and SFM data alone are also not sufficient to establish the existence of $f_0(1500)$.

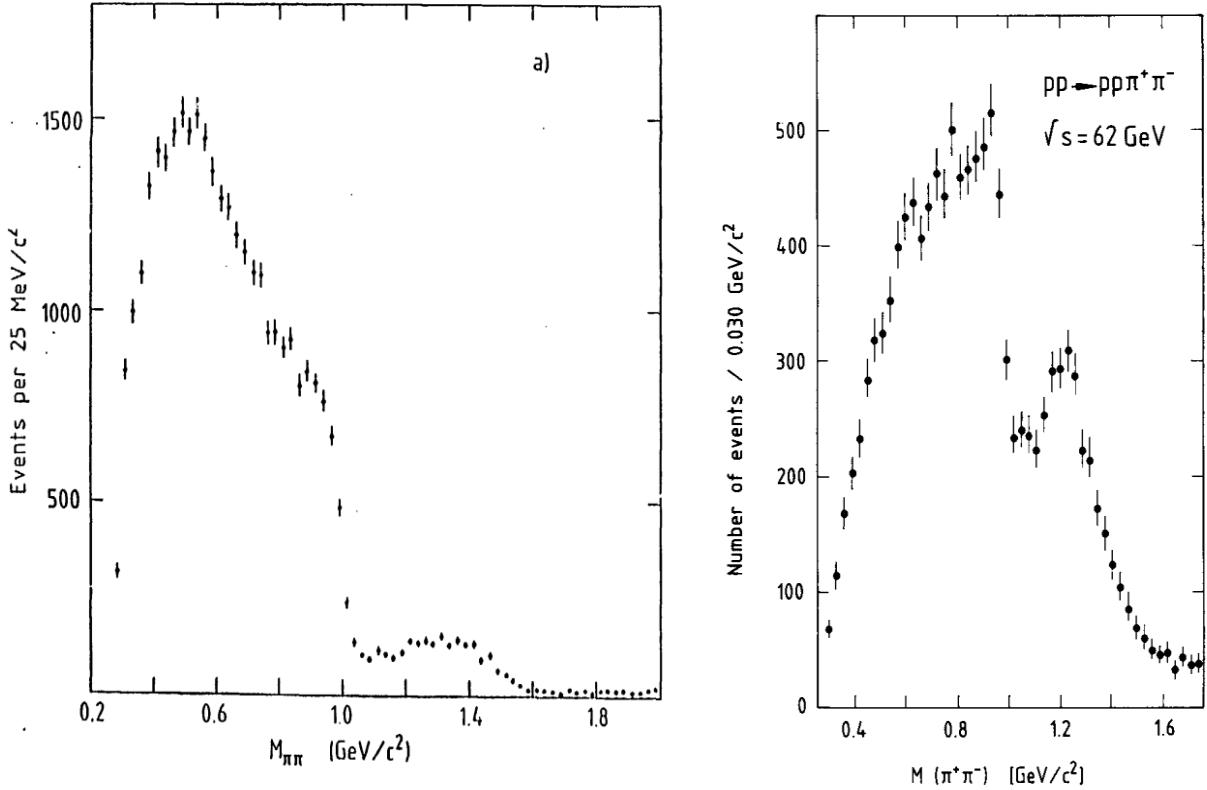

Fig. 6: $\pi^+\pi^-$ mass plot in pp CEP data at AFS (left plate, from [27]) and at SFM (right plate, from[29]).

More than 10 years later pp CEP experiments were carried out at CERN with extracted proton beams with momenta up to 450 GeV/c at the OMEGA spectrometer. Those experiments featured particle detection and identification which permitted to study with high statistics the central exclusive production channels $\pi^+\pi^-$, $\pi^0\pi^0$, $K^+K^-$, $K^0_s K^0_s$, $\eta\eta$, $\eta\eta'$ and several 4 pion channels [31-48]. For theoretical and experimental reviews see [49-55] and [56-59]. The OMEGA experiments WA76, WA91 and WA102 were performed at center of mass energies $\sqrt{s}$ =12.7, 23.2 and 29.1 GeV, lower than the 62-63 GeV of the ISR experiments. The I-tI kinematic coverage of the OMEGA experiments did not extend to as low I-tI values as in the AFS spectrometer. The $\pi^+\pi^-$ invariant mass spectrum of WA91 at $\sqrt{s}$ = 29.1 GeV (see fig 11 top plate) features a large σ signal peaking slightly above 0.5 GeV followed by a smooth descent down to about 0.9 GeV, followed by a sharp drop between 0.9 and 1.1 GeV. The signal lowers from 0.9 to 1.1 GeV by a factor less than 3. A peak is present at the mass of the $f_2(1270)$ and the signal drops from 1.1 and 1.5 GeV by a factor of about 5. The spectrum has been interpreted and well fit in the energy region 1.1-1.6 GeV as resulting from the presence of $f_2(1270)$ and of the interfering $f_0(1370)$ and $f_0(1500)$ amplitudes. The $f_0(1370)$ and $f_0(1500)$ signals are not dramatically eloquent in spite of the high statistics and, to be established, require the study of the angular distributions and delicate partial wave analysis.

The OMEGA $\pi^+\pi^-\pi^+\pi^-$ mass spectrum features a remarkable narrow peak, with width of 50-100 MeV centered at about 1450 MeV (see fig. 7 from [33]), shifted by about 50 MeV below the nominal 1.5 GeV energy of $f_0(1500)$, and too narrow to be interpreted as the $f_0(1370)$ [32-34,37]. This peak has been interpreted as the result of the interference between the $f_0(1370)$ and $f_0(1500)$ amplitudes [34]. Notice however that the $\pi^0\pi^0\pi^0\pi^0$ OMEGA pp CEP spectrum does not feature any peak around 1450 MeV (see fig. 1 in [45]). The partial wave analysis of the $4\pi$ OMEGA data privileges a solution where the $f_0(1370)$ decays into 4 pions occur dominantly via $\rho\rho$ and $f_0(1500)$ decays into 4 pions occur dominantly via $\sigma\sigma$ [45]. These results are completely at variance to the LEAR [15] and bubble chamber[18] results for the $f_0(1370)$ and $f_0(1500)$ 4 pion decays. Noticeably, however, the sum of all the fractions of $f_0(1370)$ and $f_0(1500)$ 4 pion decays of WA102 and of Crystal Barrel are compatible [15].

In view of these contradictions, under the hypothesis that the pbar data and the pp CEP data concern the same resonances, the descriptions of the $f_0(1370)$ and $f_0(1500)$ properties of pbar and CEP experiments are not compatible. Interferences between the $f_0(1370)$ (or between the broad 0.4-1.7 GeV continuum) and $f_0(1500)$ amplitudes have been introduced [34] in order to interpret the spectra and to establish mass, width and decay branching ratios of the $f_0(1370)$ and $f_0(1500)$. However for more than a decade in the PDG summary tables several papers quoted and used to give average values for mass, width and decay branching ratios of $f_0(1500)$ are discarded for the $f_0(1370)$ [6].

In the next section the main experimental lessons emerging from pp CEP data and useful for low energy spectroscopy are surveyed and used in the 4$^{th}$ section to discuss data of the STAR experiment. In our view, the $\pi^+\pi^-$ spectra of STAR pp CEP data show, among other things, that the $f_0(1370)$ is an isolated structure and not a subset of a $0^{++}$ continuum. Together with the neglected pbar data at rest, these pp CEP data provide a positive answer to the question of the independent existence of $f_0(1370)$.

**3-Central Exclusive Production experiments: expectations and experimental features**

CEP is described in terms of Regge phenomenology [49-55]. At pp colliders, when soft pp interactions generate central exclusive production the two colliding protons emerge unbroken in the final state at small scattering angles and are accompanied by a central system made of a few low energy hadrons emitted at low rapidities and separated by rapidity gaps from the two leading protons. CEP is controlled by the total cm energy $\sqrt{s}$, by the four momentum transfers I-$t_1$I and I-$t_2$I at the two proton vertices and by the angle ϕ between the scattering planes of the two protons. In CEP processes the two protons may exchange two Pomerons (PP), or one Pomeron and one Reggeon (PR), or two Reggeons (RR). These processes decrease exponentially with increasing I-tI and feature the following scaling laws with energy [49,50]:

PP ≈ s independent

PR ≈ $1/\sqrt{s}$

RR ≈ 1/s

The natural choice to single out PP interactions is then to go to high $\sqrt{s}$ center of mass energy and to as low as possible four-momentum transfer I-tI at both proton vertices. The energy depends on the accelerator for fixed target experiments and on the storage ring for collider experiments. The thickness of the liquid $H_2$ target enclosure and the threshold for detection and measurement of the spectator proton limit in fixed target experiments the smallest four momentum transfer I-$t_{min}$I which can be reached. In collider experiments the setting of the machine lattice (in practice how large can be set the lattice parameter β*) and the capability to approach the active part of the proton detectors to the stored proton beams limit I-$t_{min}$I. Elastic scattering populates the very low I-t I region and can saturate the trigger bandwidth. Nearly always and everywere during collider runs dedicated to pp elastic scattering the central detector surrounding the interaction vertex was not present or not active. One fortunate exception occurred during the P1 STAR run of 2009 at the BNL RHIC collider [60-62].

In PP exchange only $0^{++}$, $2^{++}$ etc. central systems may be produced , while also $1^{--}$, $1^{+-}$ etc. can be produced in PR and RR exchanges. ISR and fixed target experiments have provided convincing evidence of the Pomeron existence and of the s and t dependence of CEP [56-59].

AFS with pp runs at √s =45 and 62 GeV and with αα runs at √s =126 GeV has shown that the centrally produced hadronic system in CEP has properties independent on √s and on the nature of the colliding hadrons [28].

WA76 and SFM have shown substantial depression of the $ρ^0$→ $π^+π^-$ signal (which cannot be produced by PP exchange) when moving from √s = 12.7 to 21,8 GeV [57]. Moving to √s =63 GeV, the $ρ^0$→ $π^+π^-$ signal is no more visible in the $π^+π^-$ mass plot [28].

WA91 has registered a striking evidence of the advantage of measuring at low I-tI for detecting $0^{++}$ signals. WA91 compares the 4pion mass spectra measured at √s =29 GeV in the two I-tI windows (s stands for proton spectator, f stands for forward scattered proton) :

I-$t_s$I < 0.15GeV$^2$ , I-$t_f$I < 0.15GeV$^2$    and    I-$t_s$I > 0.15 GeV$^2$, I-$t_f$I > 0.15GeV$^2$.

In the low I-tI window, peaks associated to $0^{++}$ objects near 1.45 and 1.7 GeV are clearly visible besides a narrow peak due to the $f_1$(1285) meson. In the high I-tI window only the $f_1$(1285) peak survives (see fig. 7 [33]).

AFS and SFM data collected in different I-tI windows at nearly equal energies feature a $f_2$(1270) signal which is well visible in the higher I-tI window of SFM and becomes hardly appreciable in AFS at lower I-tI. Instead the σ and $f_0$(980) $0^{++}$ signals are much more relevant in the lower I-tI window (see fig- 6).

A further noticeable feature of CEP is the dependence on the angle φ between the scattering planes of the two protons and on the vector difference **d$P_t$=$P_{t1}$-$P_{t2}$** in the transverse plane of the momentum vectors of the two objects exchanged at the two proton vertices. This has been observed in SFM at ISR in 1991 as φ dependence of the $f_2$(1275) signal (breaking of vertex factorization) [30] (see fig. 8 [30]). The dependence on the angle between the scattering planes of the two protons has been observed also at OMEGA by WA91 and WA102, by measuring at √s =29 GeV the **d$P_t$=$P_{t1}$-$P_{t2}$** dependence of CEP reactions with $π^+π^-$, $K^+K^-$, and 4 pions in the central state [35,36]. At √s =13000 GeV ATLAS at LHC has observed a marked difference between $π^+π^-$ spectra collected respectively with $P_{t1} \cdot P_{t2}$ < 0 and $P_{t1} \cdot P_{t2}$ > 0 ($P_{t1}$ and $P_{t2}$ are the momentum vectors projected onto the transverse plane of the exchanged particle emitted at each proton vertex [64]. When the projections on the transverse plane of the two vectors $P_{t1}$ and $P_{t2}$ are on the same side (the relative angle is less than $90^0$), the σ and $f_0$(980) signals are enhanced and the $f_2$(1270) peak is depressed (see fig. 10 bottom plate). Conversely, when the relative angle is greater than $90^0$ the $f_2$(1270) signal is enhanced, the peak in the $f_0$(980) region is depressed and the sigma peak is suppressed.

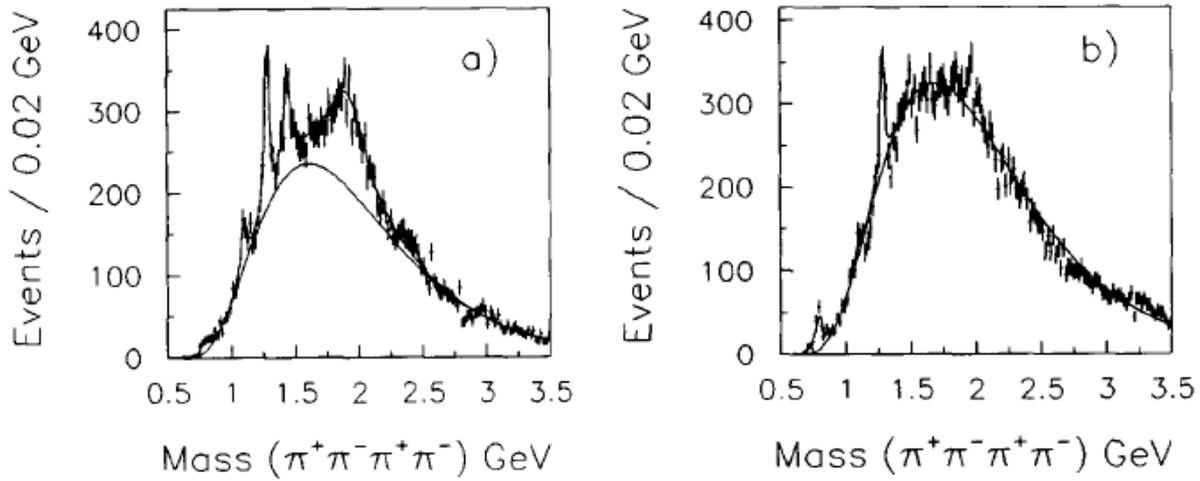

Fig. 7: π+π-π+π- mass plots measured in pp CEP by WA91 [33] for I-tI<0.15 GeV² (left) and 0.15<I-tI GeV² (right)

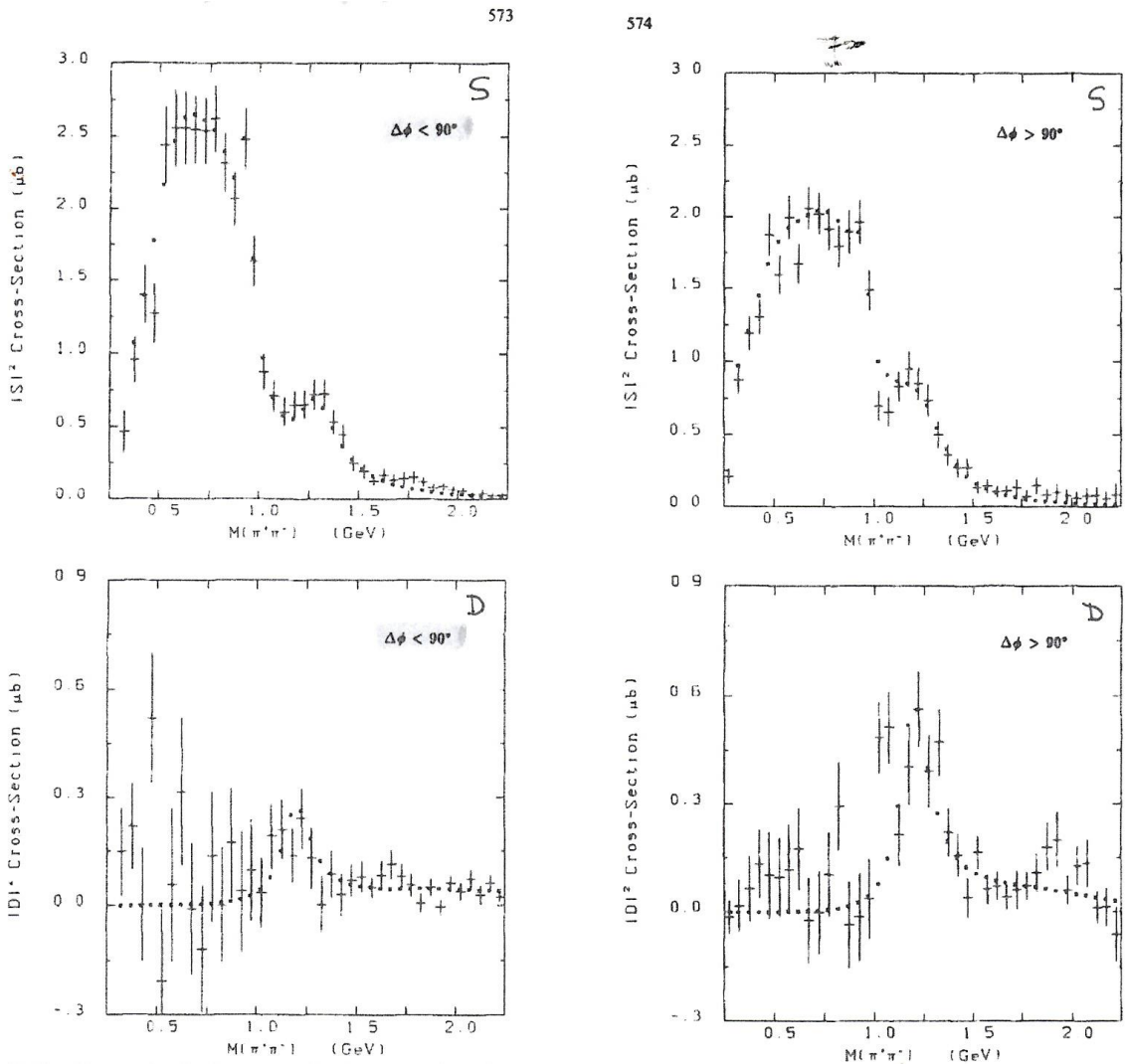

Fig. 8: Dependence on the angle φ between the scattering planes of the two protons in pp CEP measured at SFM [30]

## 4-Evidence of the existence of the $f_0(1370$ meson) in pp CEP

Fig. 9 [60] reproduces $\pi^+\pi^-$ mass plots in pp CEP data of the STAR experiment at RICH collected at $\sqrt{s}$ = 200 GeV in 2009. In the 2009 run the machine optics was optimized with $\beta^*$=20 m for a low I-tI kinematic coverage necessary for measurements of the pp elastic cross section with two sets of detectors of the forward scattered protons installed in roman pots on each side of the interaction region. The central detector which surrounds the interaction region was active and preliminary $\pi^+\pi^-$ CEP spectra were presented in 2012 and 2014 [60,61]. The spectrum discussed in [60] results from events selected in the kinematical range:

0.003 < I-$t_1$I, I-$t_2$I < 0.03 GeV$^2$  I-$t_1$I and I-$t_2$I are the four-momentum transfer to the two protons (the four momentum transfer acceptance window of these CEP data features the lowest I-$t_{min}$I of all existing CEP experiments)

I$\eta_\pi$I < 1.0  pseudorapidity of single pions,

I$\eta_{\pi^+\pi^-}$I < 2.0 pseudorapidity of the $\pi^+\pi^-$ system.

The spectrum in the left plate of fig. 9 is not corrected for acceptances and detection efficiencies. However in the selected kinematical range the acceptance of all the measured variables exceeds always 10%. The uncorrected spectrum contains 380 $\pi^+\pi^-$ events and has practically no background (the spectrum of like sign pion pairs contains few events).

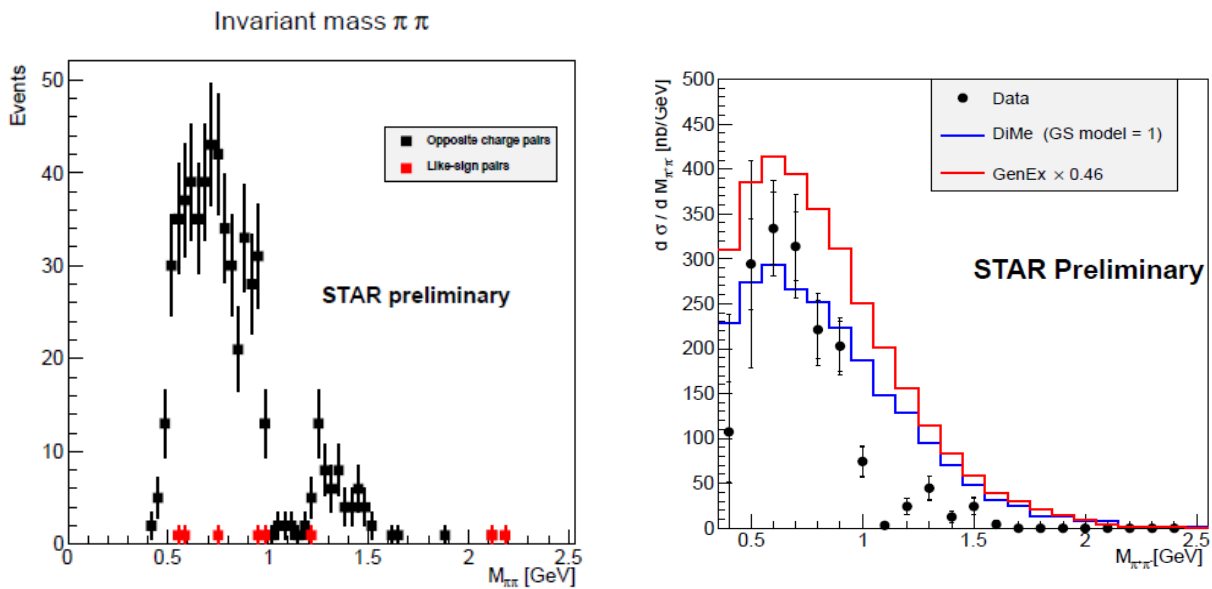

Fig. 9: $\pi^+\pi^-$ mass plot of STAR pp CEP data of the 2009 run with 0.003 < I-$t_1$I, I-$t_2$I < 0.03 GeV$^2$ kinematic coverage of both scattered protons: row data left plate [60], acceptance corrected data right plate [61].

For the 2015 STAR run the two sets of roman pots containing the proton detectors on the two sides of the central detector were approached to the interaction region and the machine optics was set with $\beta^*\approx$0.85 m in order to increase the I-tI acceptance [62]. The I-tI window at the two proton vertices was then:

0.03 < I-$t_1$I, I-$t_2$I <0. 3 GeV$^2$.

The π⁺π⁻ mass spectrum of 2015 STAR pp CEP data is reproduced in fig. 10 (2$^{nd}$ plate from top [62]).

The STAR spectrum of fig. 10 collected in the high I-tI window is similar to the π⁺π⁻ spectrum of ppbar CEP data collected at the Fermilab TEVATRON by CDF [63]. It features a signal rising from threshold to 0.9 GeV (reminiscent of the shape of the SFM data reproduced in fig. 10 top plate) and a sharp drop across the $f_0(980)$ by a factor of about 5 between the maximum at 0.9 GeV and the minimum at 1.1 GeV. A strong $f_2(1270)$ signal dominates the 1.1-1.5 GeV region. The ordinate of the valley which separates at 1.1 GeV the $f_0(980)$ drop from the $f_2(1270)$ peak is about 1/3 of the $f_2(1270)$ peak. The activity above 1.6 GeV is at a level of about 1/5 of the value of the minimum at the 1.1 GeV valley.

The STAR spectrum of fig. 9, collected in the lowest I-tI window ever explored in CEP experiments and reproduced in fig. 11 bottom plate, features a strong σ signal (reminiscent of the AFS σ signal of fig. 6) followed by a sharp drop between 0.9 and 1.1 GeV . A completely isolated structure present in the window 1.1-1.6 GeV and centered at about 1.35 GeV is generated by about 60 events. The ratio between the values of the ordinates at 0.9 and 1.1 GeV is above 15 and exceeds the value of about 10 of the AFS data [28]. The dip at 1.1 GeV is not an acceptance nor a detection efficiency effect, because the acceptance and detection efficiency vary smoothly in the full energy window relevant for σ, $f_0(980)$, $f_0(1370)$ and $f_0(1500)$. We can interpret confidently the STAR structure in the energy window 1.1-1.6 as due to dominant production of $f_0(1370)$ for several reasons: a) the contribution of $f_2(1270)$ reduces both with increasing energy and with lowering of the I-tI window; b) the S-wave contribution remains constant with increasing $\sqrt{s}$ and therefore the relative s-wave contribution grows with increasing energy (already at √ s =63 GeV the S-wave dominates the 1.1-1.6 energy region [28]), c) the width of the structure is too large to be associated to the $f_0(1500)$ and extends down to 1.2 GeV. The statistics is too low to extract a ratio between the $f_0(1370)$ and $f_0(1500)$ contributions, however there are more events at masses below 1.4 GeV.

The S-wave signal below the $f_2(1270)$ peak in the ATLAS-ALFA 13 TeV data [64] produced in a high I-tI window (see fig 10 bottom plate) generates a shoulder on the left of 1.5 GeV. This suggests that the contribution of the $f_0(1500)$ to π⁺π⁻ CEP is below that of $f_0(1370)$ and that it could manifest as a destructive interference with the high energy tail of $f_0(1370)$ depressing the π⁺π⁻ spectrum in the 1.5 GeV region.

Independently of the interpretation of the structure in the 1.1-1.6 GeV energy region, quite noticeable is the fact that the STAR π⁺π⁻ spectrum drops nearly to zero at 1.1 GeV. This may be the result of the interference of the amplitudes of the low energy tail of $f_0(1370)$ with the high energy tail of $f_0(980)$ plus the effect of the vicinity of the KKbar threshold, but very likely it is due to absence of the S-wave continuum. The $0^{++}$ continuum, which is usually invoked with its destructive interference with the $f_0(980)$ amplitude to generate the drop around 1 GeV, seems drastically reduced. It looks like the σ meson which generates the broad peak above 0.5 GeV is confined below 1 GeV. In other words it looks like the red-dragon proposed by Minkowsky and Ochs [1], that featured a low energy body centered at about 0.6 GeV and a head extending below the $f_0(1500)$, is split into two separate parts, the σ and the relatively narrow $f_0(1370)$ (which was visible since long ago in pbar annihilations at rest).

**5- Summary, conclusions and prospects**

The π⁺π⁻ mass spectra in pp CEP data collected in high I-tI windows by experiments at ISR [29] and at high energy colliders STAR [62], CDF [63] and ATLAS [64] (see fig. 10) )have similar features: flat or rising slope before 0.9 GeV, a sudden drop around the $f_0(980)$ occurring in a window between 0.9 and 1.1 GeV, a sharp peak between 1.1 and 1.5 GeV dominated by $f_2(1270)$, a tail with gentle slope from 1.5 to and beyond 2 GeV.

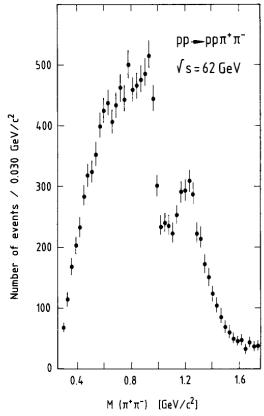

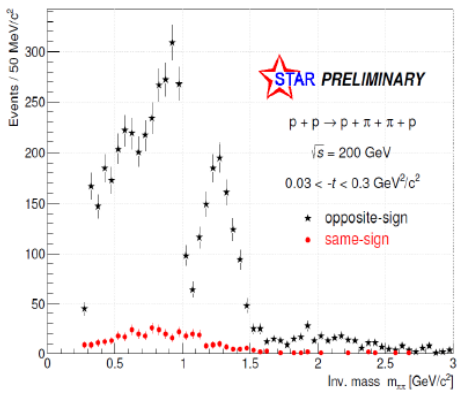

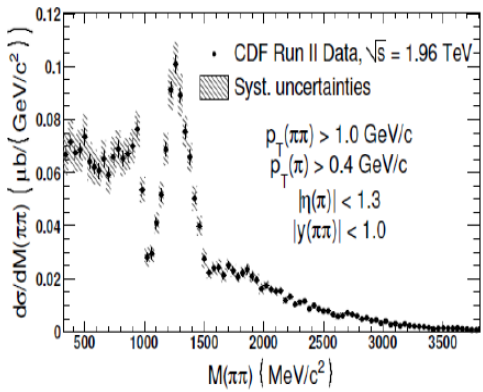

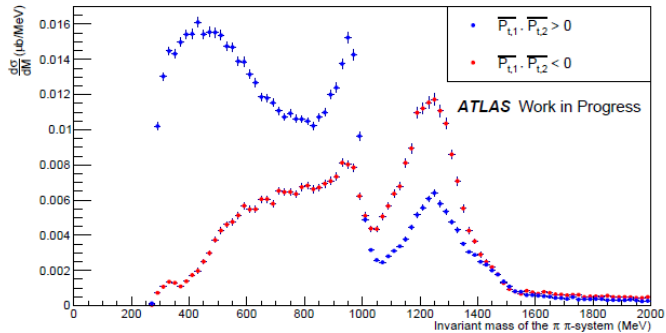

Fig. 10: $\pi^+\pi^-$ mass plots of pp CEP data collected by SFM [29], STAR [62], CDF [63] and ALPI-ATLAS [64] at increasing cm energies $\sqrt{s}$ =62, 200, 1960 and 13000 GeV in high I-tI windows.

The $\pi^+\pi^-$ mass spectra in pp CEP data collected in low |-t| windows by AFS [27,28] and STAR [60,61] see fig. 11 have in common:

a descending slope from about 0.6 GeV to 0.9 GeV followed by

a sharp drop from about 0.9 to 1.1 GeV around the $f_0(980)$ mass

a S-wave dominated bump between 1.1 and 1.6 GeV

The |-t| region covered by STAR in 2009 extends to lower |-t| values than in AFS. We expect then in the STAR 2009 data an S-wave contribution more pure than in the AFS data.

The ordinate at the intermediate minimum at 1.1 GeV in the AFS $\pi^+\pi^-$ spectrum is ten times larger than the minimum at 1.6 GeV before the gentle rise towards 2 GeV. In the low l-tl window 2009 STAR data the ordinates at the 1.1 and at 1.6 GeV minima have very low values. This may simply be because of the nearly complete absence of $f_2(1270)$ and of S-wave continuum in data in the very low l-tl region at high energies.

The spectacular increase of $\sqrt{s}$ moving to the RICH, then to the Tevatron and then to the LHC collider helps a lot in isolating Pomeron-Pomeron contributions in CEP interactions, since Pomeron-Reggeon and Reggeon-Reggeon contributions are respectively suppressed according to the $1/\sqrt{s}$ and $1/s$ dependence on s.

Pomeron-Pomeron interactions can produce, besides $0^{++}$, also $2^{++}$ etc. objects with zero isospin but, by lowering the l-tl window, the $2^{++}$ contribution is reduced, as witnessed by the $f_2(1270)$ production dependence on t at ISR (see fig. 6).

Data on $K^+K^-$, $K^0_sK^0_s$, $\pi^+\pi^-\pi^+\pi^-$ CEP production at very high energy and at low I-tI are not yet available to complete the analysis of this paper.

The $\pi^+\pi^-$ mass plot of STAR pp CEP data feature an isolated $f_0(1370)$ peak. Bubble chamber data of npbar annihilations at rest feature a clear peak of $f_0(1370)$ decaying into $K^0_sK^0_s$; the $K^0_sK^0_s$ band in the $K^0_sK^0_s\pi^-$ Dalitz plot intercepts and interferes with the $K^{*-}$ bands. These facts support firmly the interpretation of the $f_0(1370)$ as an isolated resonance worth being treated as a bona fide $0^{++}$meson. It is also worth mentioning that clear $f_0(1370)$ signals are visible in radiative decays of $J/\psi$ and of $\psi(2S)$ measured at CLEO [65] on the right of the $f_2(1270)$ peak in $\gamma\pi^+\pi^-$ and $\gamma\pi^0\pi^0$ decays and on the left of the $f_2'(1525)$ peak in $\gamma K^+K^-$ and $\gamma K^0_sK^0_s$ decays (see fig. 9 of ref. [65]).

The hierarchy of the decay branching ratios BR of the $f_0$ mesons: BR($\sigma\sigma$), BR($\rho\rho$), BR($\pi^*\pi$), BR($\pi\pi$), BR(KKbar) etc. depends on the qqbar and gg composition of the $f_0$ physical states and of the couple of mesons into which they decay [70,71] and on the possible violation of flavor democracy of gg decays [72].

Work is still necessary to measure decays into $K^+K^-$ and $K^0_sK^0_s$ of $f_0(1370)$ produced in pp CEP at high energies at low I-tI and to compare them with $\pi^+\pi^-$ decays of $f_0(1370)$ produced in the same conditions of kinematical selections, in order to establish the ratios between the decay branching ratios BR($K^+K^-$)/BR($\pi^+\pi^-$) and BR($K^0_sK^0_s$)/ BR($\pi^+\pi^-$).

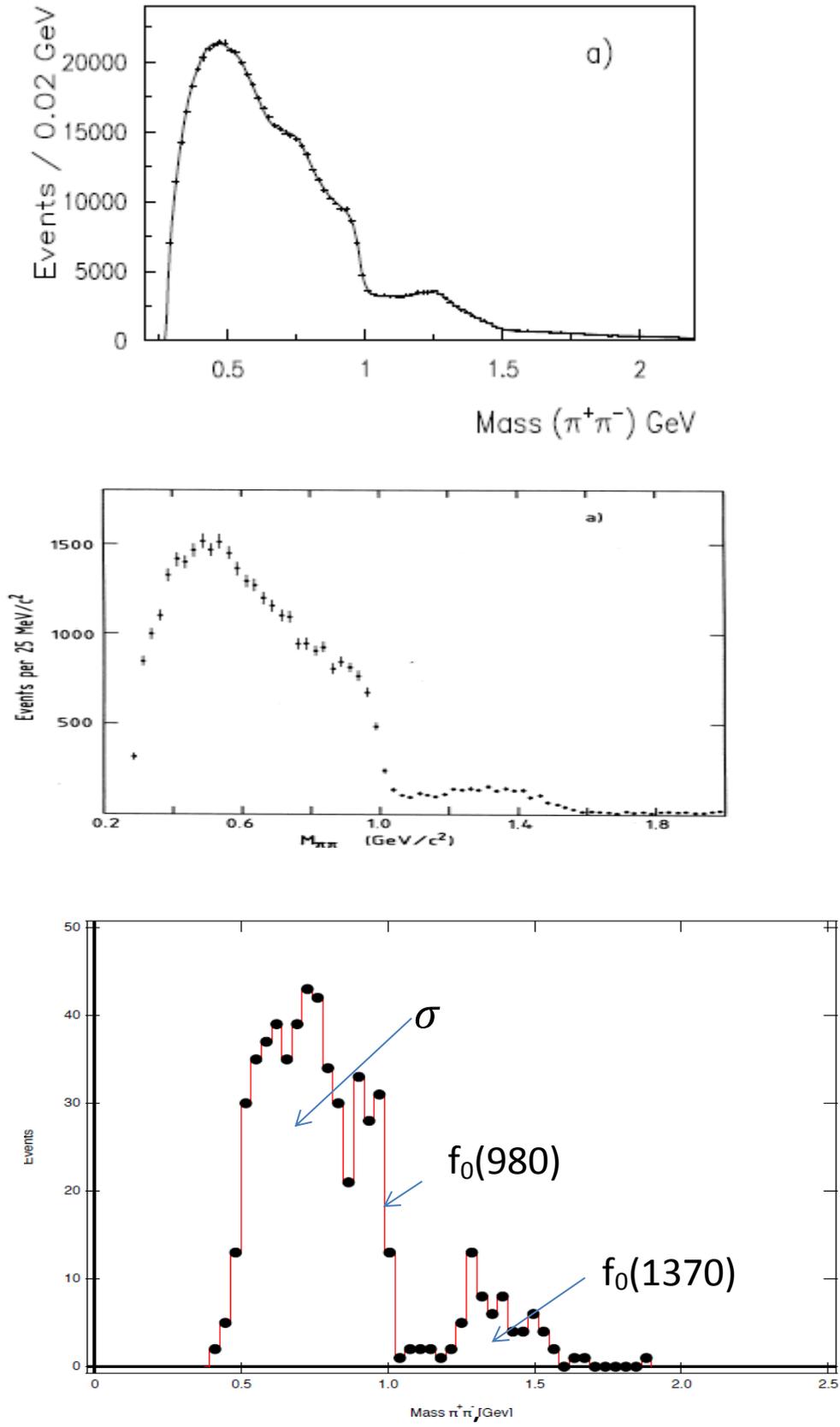

Fig. 11 : $\pi^+\pi^-$ mass plots of pp CEP data collected in low |-t| windows respectively (from top to bottom) by WA91 [34], AFS [27] and STAR (data points from [60]) at increasing cm energies $\sqrt{s}$ = 29, 63 and 200 GeV.

Work is also necessary to investigate $f_0(1370)$ decays into 4 charged pions produced in pp CEP at low |-t|, in order to reduce or eliminate non-PP contributions to the spectrum, understand the narrow peak at 1450 MeV present in WA76 [32], WA91 [33,34] and WA102 [37] data and establish the ratios of the decays into 4 charged pions via σσ and ρρ of $f_0(1370)$ and $f_0(1500)$ produced in pp CEP.

As mentioned in the introduction, there are possibilities that σ and $f_0(980)$ have a non qqbar nature [1-13, 73-76]. Under this circumstance one candidate among $f_0(1370)$, $f_0(1500)$ and $f_0(1710)$ would be in excess for the two isoscalar places in the $0^{++}$ nonet of qqbar mesons. In order to play the exercise of unfolding the possible mixing between the qqbar and gg components of the three $f_0$ physical mesons their relative decay branching ratios into pairs of pseudoscalar mesons need to be revisited critically, in particular for KKbar decays, both for pbar annihilation data and CEP data. All the three observed mesons have mass not far from predictions of lattice QCD calculations (see [66-69] and refs. therein). QCD sum rules and low energy theorems [73-76] predict dominant σσ and ηη' decays for gluonium in the 1-2 GeV mass window. $f_0(1370)$ has too low mass to decay into ηη'. In view of the dominant σσ decays of $f_0(1370)$ measured in pbar annihilations at rest, it is then not impossible that gluonium has been with us since bubble chamber times of the sixties [17,18,23-25] before the invention of QCD.

Moving to LHC CEP data, we expect that by cumulating the effects of increasing √s and of reducing |-$t_{min}$| and selecting |-$t_1$|, |-$t_2$| in windows near |-$t_{min}$|, nearly pure S-wave production will be selected and that the continuum be dramatically reduced. This should remove from the $π^+π^-$ mass spectra the $f_2(1270)$ and from the $K^+K^-$ mass spectra the $f_2'(1525)$ and clean up the 1.1-1.6 GeV energy region leaving there only the contributions of the interfering amplitudes of $f_0(1370)$ and $f_0(1500)$.

In available CEP data the contribution of $f_0(1500)$ in the $π^+π^-$ channel looks marginal and noticeable at most as generating a destructive interference, which creates a descending slope from 1.4 to 1.6 GeV, while in the $K^+K^-$ channel the $f_0(1500)$ signal is obscured in the mass plot by the presence of the $f_2'(1525)$.

With new LHC data it should be possible to establish with confidence the relative decay branching ratios into $π^+π^-$, $K^+K^-$ and $K^o_s K^o_s$ of $f_0(1370)$ and $f_0(1500)$ in CEP production and to compare them with the values measured in pbar annihilations at rest.

Another major task will be the study of the signal due to the interfering amplitudes of $f_0(1370)$ and $f_0(1500)$ in the $π^+π^- π^+π^-$ decay channel. In dpbar annihilations at rest the $f_0(1370) → π^+π^- π^+π^-$ decays nearly saturate the production of the 5 pion final state. It remains to be seen what will happen in $0^{++}$ $π^+π^-π^+π^-$ CEP production, where the most remarkable feature observed at the OMEGA spectrometer is the rather narrow peak in the $π^+π^- π^+π^-$ mass spectrum at about 1450 MeV [32-34,37] which has been interpreted [34,37] as the result of the interference of $f_0(1370)$ and $f_0(1500)$ amplitudes.

2015 data with not too low statistics have been collected at LHC by the CMS-TOTEM and ATLAS-ALFA Collaborations and should be usable to check our qualitative conclusions. The 2015 data have been collected with β*=90 m optics of LHC. The $t_{min}$ covered may extend down to 0.04 $GeV^2$ [77]. Other special optics runs where ATLAS-ALFA and CMS with upgraded TOTEM detectors [78] will collect data are foreseen in 2018.

With larger β* as used for measurements at LHC of pp elastic scattering cross sections, the |-t| range could be extended below 0.003 $Gev^2$ and CEP could be measured in kinematical regimes where Pomeron-Pomeron interactions should be definitively dominant and $0^{++}$ production overwhelming. All $0^{++}$ low energy

structures should emerge in the mass spectra over a reduced or missing S-wave continuum. A detailed study of σ, $f_0(980)$, $f_0(1370)$, $f_0(1500)$, $f_0(1710)$ and of what will eventually remain of the low energy scalar continuum should become feasible. If gluonium, predicted more than 40 years ago [79,80], might represent for non-perturbative QCD something like the hydrogen atom or positronium represented for QED, it might well be worth using LHC and its detectors during a while as a low mass $0^{++}$ factory.